\documentclass[pra,twocolumn,showkeys,showpacs,superscriptaddress,floatfix,longbibliography]{revtex4-1}

\usepackage[T1]{fontenc}
\usepackage{epsfig,psfrag}
\usepackage{latexsym}
\usepackage{graphicx}
\usepackage{dcolumn}
\usepackage{amssymb}
\usepackage{amsmath}
\usepackage{bm}
\usepackage{mathrsfs} 
\usepackage{bigints}
\usepackage{upgreek}
\usepackage{color}
\usepackage{longtable}
\usepackage{xspace}
\usepackage[squaren]{SIunits} 
\usepackage{epstopdf}
\usepackage{float}
\usepackage[normalem]{ulem}
\usepackage{hyperref}
\usepackage{bbold}
\usepackage{bbm}
\usepackage{physics}
\definecolor{mygreen}{RGB}{20,148,20}

\usepackage{filemod}

\newcommand{\eq}[1]{Eq.~\eqref{eq:#1}}

\newcommand{\fig}[1]{Fig.~\ref{fig:#1}} 

\newcommand{\sect}[1]{Sec.~\ref{sec:#1}}

\newcommand{\gstate}{5s^2\,^1\textrm{S}_0}
\newcommand{\OnePOne}{5s5p\,^1\textrm{P}_1}
\newcommand{\ThreePZero}{5s5p\,^3\textrm{P}_0}
\newcommand{\ThreePOne}{5s5p\,^3\textrm{P}_1}
\newcommand{\ThreePTwo}{5s5p\,^3\textrm{P}_2}
\newcommand{\PTwo}{5p^2\,^3\textrm{P}_1}
\newcommand{\ThreeDOne}{5s5d\,^3\textrm{D}_1}

\DeclareUnicodeCharacter{2212}{-}

\begin{document}

\title{Magic wavelength at 477 nm for the strontium clock transition} 

\author{Xinyuan Ma}
\affiliation{Centre for Quantum Technologies, National University of Singapore, 117543 Singapore, Singapore}
\author{Swarup Das}
\affiliation{Nanyang Quantum Hub, School of Physical and Mathematical Sciences, Nanyang Technological University, 21 Nanyang Link, Singapore 637371, Singapore}
\affiliation{MajuLab, International Joint Research Unit IRL 3654, CNRS, Universit\'e C\^ote d'Azur, Sorbonne Universit\'e, National University of Singapore, Nanyang
Technological University, Singapore}
\author{David Wilkowski}
\email{david.wilkowski@ntu.edu.sg}
\affiliation{Centre for Quantum Technologies, National University of Singapore, 117543 Singapore, Singapore}
\affiliation{Nanyang Quantum Hub, School of Physical and Mathematical Sciences, Nanyang Technological University, 21 Nanyang Link, Singapore 637371, Singapore}
\affiliation{MajuLab, International Joint Research Unit IRL 3654, CNRS, Universit\'e C\^ote d'Azur, Sorbonne Universit\'e, National University of Singapore, Nanyang
Technological University, Singapore}
\author{Chang Chi Kwong}
\affiliation{Nanyang Quantum Hub, School of Physical and Mathematical Sciences, Nanyang Technological University, 21 Nanyang Link, Singapore 637371, Singapore}
\affiliation{MajuLab, International Joint Research Unit IRL 3654, CNRS, Universit\'e C\^ote d'Azur, Sorbonne Universit\'e, National University of Singapore, Nanyang
Technological University, Singapore}


\begin{abstract} 
We report the experimental measurement of a magic wavelength at 476.82362(8)~nm for the $^{88}$Sr $\gstate\rightarrow\ThreePZero$ clock transition. The magic wavelength is determined through an AC-Stark shift spectroscopy of atoms in an optical dipole trap. The value is in reasonable agreement with the theoretical prediction, differing by 0.061(54) nm. This magic wavelength, being shorter than the commonly used one at 813~nm, will be important for applications such as Bragg-pulse matter-wave interferometry involving both clock states. This work also paves the way for quantum simulation with a shorter lattice constant. 

\end{abstract}

\maketitle 

\section{Background}

Laser-cooled neutral atoms are studied for various applications, such as quantum simulation and computing~\cite{Bloch2012,Henriet2020quantumcomputing}, precision sensing~\cite{geiger2020}, and metrology~\cite{ludlow2015}. These applications usually require optical trapping of atoms and coherent control of a transition between two internal states under consideration. Common trapping techniques include large optical dipole traps, optical lattices, and optical tweezers. With sufficiently deep optical lattices and tweezers, it is possible to suppress Doppler broadening and recoil shifts, allowing for high-precision spectroscopy. However, within the trap, spatial inhomogeneity generally reduces the coherence between the states. A common technique to suppress this dephasing mechanism consists in tuning the optical lattice wavelength to the magic wavelength, canceling the differential AC-Stark shift~\cite{Katori2003,Takamoto2003,Ye2008}. 

The magic wavelength value depends on the atomic species and the transition under consideration. Ultra-narrow clock transitions of alkaline-earth atoms have attracted a lot of interest for quantum technological and precision measurement applications. Optical lattice clocks~\cite{Takamoto2005,Lemke2009,Bothwell2019,Aeppli2024} that operate on these transitions are among the most precise and accurate clocks and could play an important role in a future redefinition of the second~\cite{Dimarcq2024}. For the $\gstate\rightarrow\ThreePZero$ clock transition of strontium atoms at 698~nm, the optical lattice generally operates at a magic wavelength of 813~nm~\cite{Takamoto2003}. 813~nm magic-wavelength laser beams have also been used in strontium tweezer array systems~\cite{Covey2019, Schine2022} and quantum gas microscopy~\cite{Buob2024}, contributing to applications of the clock transition in quantum computation~\cite{Tsai2025} and quantum simulation~\cite{Kolkowitz2017}.

For a given transition, there could be more than one magic wavelength. Finding new magic wavelengths can facilitate some specific usages. For example, the recent measurement of a magic wavelength close to 461~nm for the strontium $\gstate\rightarrow\ThreePOne$ intercombination line~\cite{Kestler2022} led to progress in the state-insensitive trapping of atoms close to a nanofiber~\cite{Kestler2023}.  For the strontium clock transition, we expect three magic wavelengths that are red detuned from the main $\gstate\rightarrow\OnePOne$ transition at 461~nm, namely with a positive real part of the polarizability for each clock state. So far, only the magic wavelength at 813~nm has been used, mainly because the photon scattering rate is low, leading to potentially long coherence times. Recently, a shorter magic wavelength at 497 nm has been identified \cite{kestler2025turquoise}. Shorter wavelengths can be useful to create an optical lattice with a smaller inter-trap spacing. In addition, magic wavelength Bragg pulses that are used in matter-wave interferometers operating on both clock states~\cite{Li2023}, can benefit from shorter wavelengths that transfer larger momenta to the atoms.

\begin{figure}[t!]
    \centering
    \includegraphics[width=\linewidth]{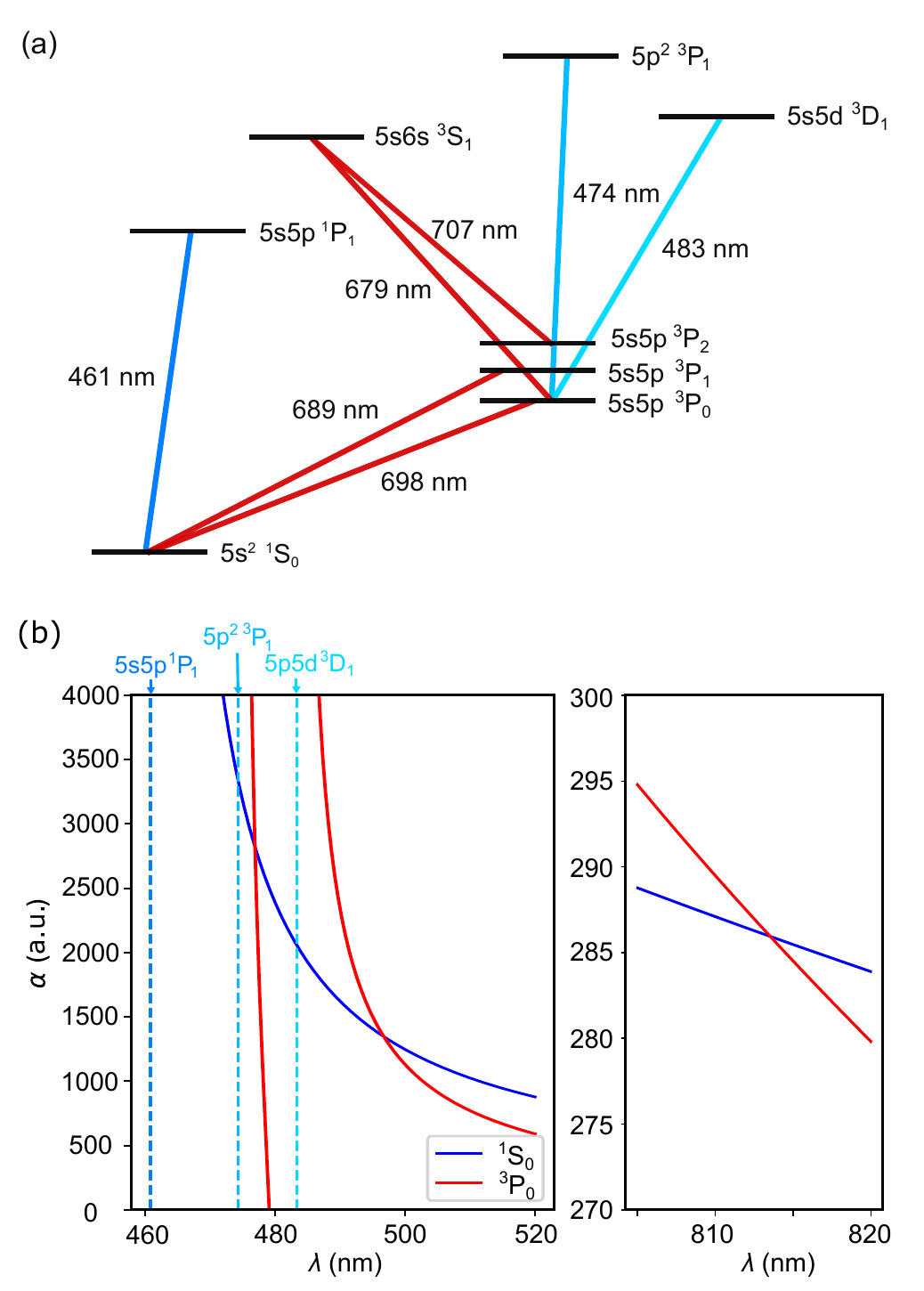}
   \caption{(a) The Sr energy levels and the transitions relevant to this study (not drawn on scale). The transition at 461~nm is used for the first-stage MOT and the imaging of the ground state. The transitions at 679~nm and 707~nm are used to repump the atoms from the $\ThreePZero$ and $\ThreePTwo$ metastable states during the first-stage MOT. They are also used to transfer the $\ThreePZero$ population to the $\gstate$ state for fluorescence imaging. The 689~nm transition is used for the second-stage MOT. The 698~nm transition is the clock transition for which the magic wavelength is measured. (b) Wavelength dependences of the $\gstate$ (blue curve) and $\ThreePZero$ (red curve) clock state polarizabilities in atomic unit (a.u.) plotted in the blue (left panel) and the near-infrared (right panel) spectral regions. The crossing points of the polarizability curves give the expected values of the red-detuned (\textit{e.g.} $\alpha>0$) magic wavelengths. We find three magic wavelengths at 477~nm, 497~nm, and 813~nm. The first one is the focus of this study. The cyan-dashed vertical lines on the left panels indicate the position of the 474~nm and 483~nm resonances between the metastable $\ThreePZero$ states and the excited states indicated at the top of the figure. These transitions give the main contribution to the $\ThreePZero$ polarizability at 477 nm. In addition, the 461~nm $\gstate\rightarrow\OnePOne$ transition (blue-dashed line) gives the dominant contribution to the ground-state polarizability, as indicated in Table \ref{tab:dominant_transitions}.}
    \label{fig:Intro}
\end{figure}

In this work, we experimentally measure the shortest red-detuned magic wavelength for the strontium clock transition, using the bosonic $^{88}$Sr isotope. This magic wavelength, which is at approximately 477~nm, lies between the 474~nm  $\ThreePZero\rightarrow\PTwo$ transition and the 483~nm $\ThreePZero\rightarrow\ThreeDOne$ transition [see~\fig{Intro}(a) for the energy level diagram of strontium]. As the closest red-detuned magic wavelength to the $\gstate\rightarrow\OnePOne$ 461~nm transition, the atomic polarizability is the largest among the red-detuned magic wavelengths, as shown in \fig{Intro}(b). The magic wavelength is determined by spectroscopic measurements of the frequency shift of the 698~nm clock transition in the presence of an optical dipole trap. Then, we compare the measured magic wavelength with a theoretical estimation. 

This article is organized as follows. In \sect{theo}, we present the theoretical estimation of the magic wavelength. In \sect{expt}, we discuss the experimental setup and the procedure for the spectroscopic measurements. In \sect{results}, we report the results on the 477~nm magic-wavelength measurement. In addition, we compare the differential AC-Stark shift measurements with a model that includes the residual inhomogeneous broadening of the clock transition line. In \sect{conclusion}, we summarize our studies and provide an outlook for the uses of the 477-nm magic wavelength.

\section{Polarizabilities estimation} \label{sec:theo}

The AC-Stark shift of a state $i$ in the electric dipole approximation is given by
\begin{equation}
  \Delta E_i = -\frac{1}{2} \alpha_i(\omega_L,\epsilon_L) E_L^2,
\end{equation}
 where the state polarizability $\alpha_i$ depends on the laser frequency $\omega_L$ and polarization $\epsilon_L$. $E_L$ is the electric field amplitude of the laser.

\begin{table}[t]
\centering
 \begin{tabular}{||c c c c c 
 ||} 
 \hline
 Contribution & & $\lambda\,(\textrm{nm})$& $D\,(\textrm{a.u.})$& $\alpha_0\,(\textrm{a.u.})$ \\ [0.5ex] 
 \hline\hline
 $\ThreePZero\rightarrow\ThreeDOne$ &  &483.3486&2.450(24)&-1555(31) \\ 
 $\ThreePZero\rightarrow\PTwo$ &  &474.3157&2.605(26)&4382(88) \\ 
Other &  &  & &-7(1) \\
Total &  &  & &2820(94) \\
  & & & &\\
$\gstate\rightarrow\OnePOne$ &  &460.8719&5.248(2)&2813(3) \\
Other &  &  & &7(1) \\
Total &  &  & &2820(4) \\
 \hline
 \end{tabular}
\caption{List of different transition wavelengths in vacuum and recommended dipole matrix elements in atomic units (a.u.) for the $\ThreePZero$ and $\gstate$ states. The last column gives a breakdown of all the contributions to the polarizability $\alpha_0$ at 476.885 nm. The values are extracted from \cite{Safronova2013}.   }\label{tab:dominant_transitions}
\end{table}

\begin{figure}[h!]
\centering
  \centering
  \includegraphics[width=1\linewidth]{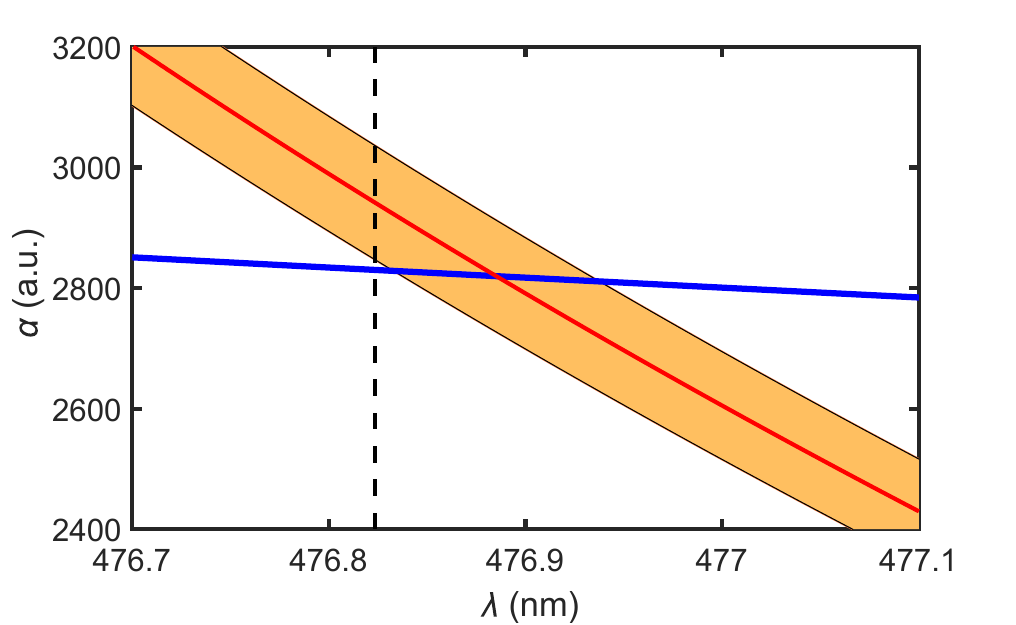}
\caption{The vertical dashed line indicates the position of the 477-nm magic wavelength found in the experiment. The blue (red) curve represents the polarizability of the $\gstate$ ($\ThreePZero$) state calculated using the values tabulated in Table~\ref{tab:dominant_transitions}. The orange-shaded area represents the uncertainty in the polarizability of the $\ThreePZero$ state due to the uncertainties in the dipole matrix elements. The uncertainty in the $\gstate$ state polarizability is comparable to the thickness of the line. The uncertainty in the experimental measurement of the magic wavelength is too small to be visible on this scale.}
  \label{fig:freq_pol}
\end{figure}

Generally, the polarizability can be decomposed into three contributions: the scalar, the vector, and the tensor polarizabilities. However, due to the spherical symmetry of the bosonic strontium clock states, the vector and tensor terms vanish. Thus, the polarizability for state $i$ with total angular momentum $J=0$ takes the following polarization-independent form~\cite{Mitroy2010}, 
\begin{equation}
 \alpha_i(\omega_L) = \frac{4\pi}{3h}\sum_n\frac{\omega_{ni}}{\omega_{ni}^2-\omega_L^2} | \bra {i}|D|\ket{n}|^2,   
 \label{eq:polarizability}
\end{equation}
where $\omega_{ni}$ is the angular frequency of the bare transition from state $i$ to state $n$, $D$ is the electric dipole operator, and $h$ is the Planck constant. The equation clearly indicates that the transitions with nearby resonant frequencies present the dominant contribution. These transitions are shown in \fig{Intro}(a) and their respective properties are listed in Table~\ref{tab:dominant_transitions}. Using \eq{polarizability} and the values given in Table~\ref{tab:dominant_transitions}, we compute the polarizabilities of the clock states around 477 nm, as shown in Fig. \ref{fig:freq_pol}.

\section{Experiment}\label{sec:expt}

The experiment is performed on an ultracold cloud of $^{88}$Sr bosonic isotope. A schematic drawing of the relevant transitions is shown in \fig{Intro}(a). Details of the experimental setup have been described in Ref.~\cite{Li2022}. In short, the $^{88}$Sr atomic ensemble is produced in a two-stage magnetooptical trap (MOT). The dipole-allowed $\gstate\rightarrow\OnePOne$ transition at 461~nm and the intercombination $\gstate \rightarrow\ThreePOne$ transition at 689~nm are used for the first and second cooling stages, respectively. At the end of the cooling stages, we have $\sim 10^7$ atoms at a temperature of $1.0(1)$~$\mu$K. 

Then, the atoms are transferred into a three-dimensional optical lattice operating at the 813-nm magic wavelength. The optical lattice is generated by a 4.8~W laser (Precilasers system) that is divided into three mutually orthogonal retro-reflected beams that overlap at the atomic cloud. Each lattice beam has a waist of $w_T=90(3)\,\mu$m, and a power of 650(5) mW. The mean trapping frequency is measured to be  $55$~kHz. To ensure that the lattice beams do not interfere with each other, a frequency difference of around $1\,$MHz, much larger than the trapping frequency is kept among the beams along the three directions. The atoms are confined in the Lamb-Dicke regime ($\eta\sim$\ 0.28). $N_0=1.0(1)\times10^6$ atoms within a cloud of horizontal size $\sigma_{x,y} =60(4)$~$\mu$m and vertical size $\sigma_z = 35(2)$~$\mu$m are loaded into the lattice.  

\begin{figure}
    \centering    \includegraphics[width=\linewidth]{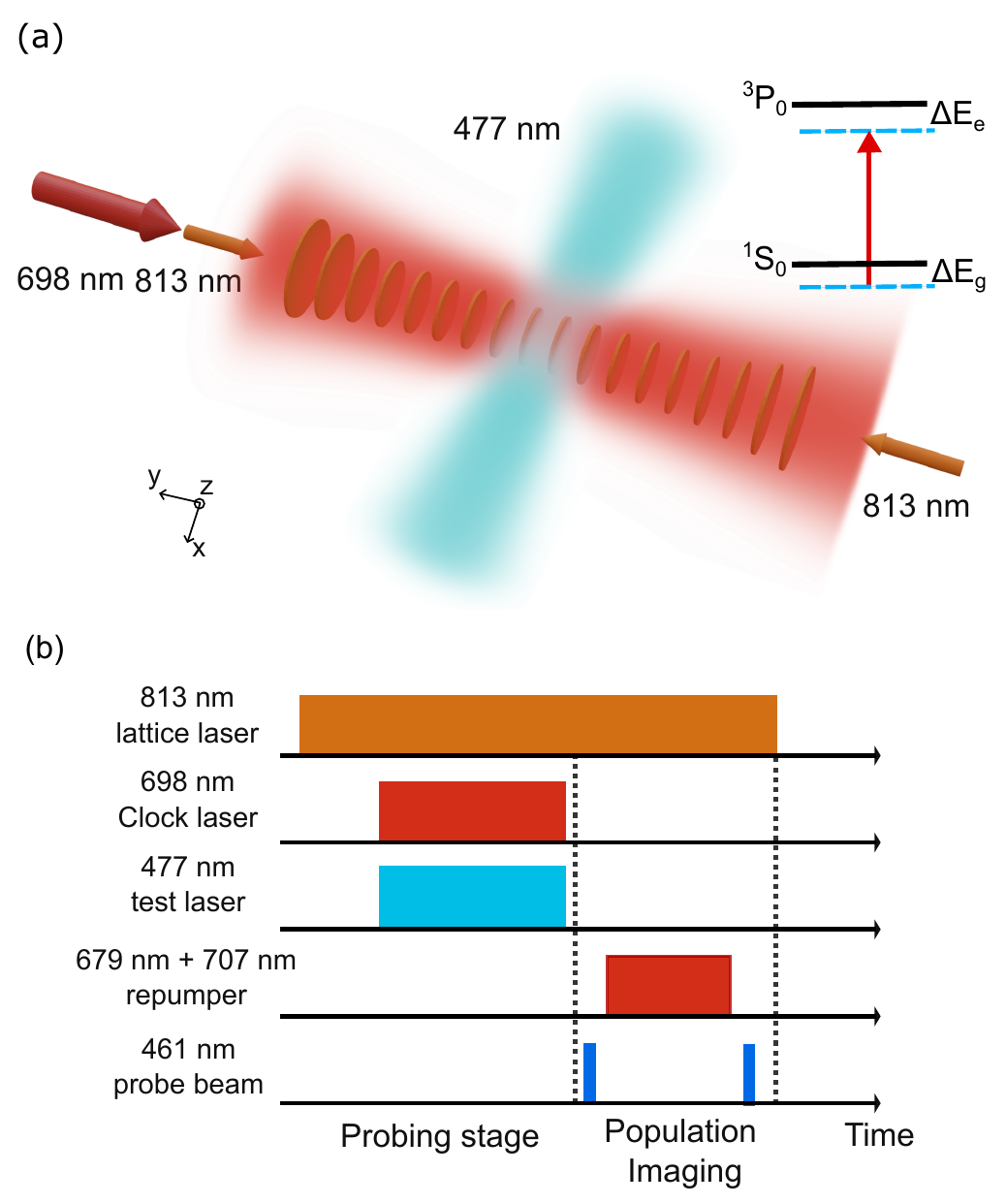}
    \caption{(a) Schematic of the laser beams involved in the AC-Stark shift spectroscopy. The 698~nm running wave interrogates the atoms on the clock transition. The 813~nm lattice beams are used to trap atoms and to suppress the Doppler and recoil frequency shifts (Lamb-Dicke regime). For the sake of simplicity, only the direction of the three-dimensional optical lattice beam along the clock beam is represented. The 477~nm beam is sent along a direction that is approximately perpendicular to the clock laser beam. The ground and excited state experience AC-Stark shifts due to this 477-nm beam by an amount $\Delta E_{g}$ and $\Delta E_{e}$, respectively. This is schematically represented by blue-dashed lines in the inset, whereas the red arrow indicates the frequency-shifted clock transition. (b) Experimental timing sequence. For representation purposes, the duration of laser beams and the gaps between operations are not drawn in proportion. The 477~nm is turned on and off in an interleaved manner over the experimental runs. The first (second) 461~nm pulse is used for ground (excited) state population measurement.}
    \label{fig:expt}
\end{figure}

A clock laser beam at 698 nm, derived from an external cavity diode laser that is stabilized on an ultra-low expansion cavity (Menlo system), is overlapped with one of the horizontal optical lattice beams using a dichroic mirror [see~\fig{expt}(a)]. The linewidth of the clock laser beam, after frequency stabilization and linewidth narrowing on the cavity, is below 10~Hz. The strontium clock transition is doubly forbidden, and in the case of the fermionic isotope, it is weakly allowed due to state mixing arising from hyperfine interactions~\cite{Boyd2007}. As $^{88}$Sr has zero nuclear spin, we applied a static magnetic field of 60~G along the polarization axis of the clock laser to activate the clock transition through a state mixing between the $\ThreePZero$ and $\ThreePOne$ states~\cite{Taichenachev2006}. The clock laser beam waist is 360~$\mu$m, making it at least three times larger than the cloud size to minimize inhomogeneous dephasing that might broaden the transition. During the clock-transition interrogation, we maintain the power of the clock laser at 5~mW, leading to a Rabi frequency of about $25\,$Hz. In addition, the clock transition experiences a global (quadratic Zeeman and AC-Stark shifts at 698 nm) frequency shift of about $1\,$kHz, giving a negligible shift of the magic wavelength position. 

A frequency-tunable 140~mW diode laser at 477~nm (Toptica DL pro) is used as a test laser. The beam has a weak ellipticity leading to $(w_x,w_z)=(160,170)\,\mu$m, where $w_x$ ($w_z$) is the waist along the $x(z)$-axis. We stabilize the laser frequency on a wavemeter (HighFinesse WS7-30) through its software-based PID controller. To guarantee its ultimate 30~MHz accuracy, the wavemeter is referenced on the 698~nm clock-transition laser~\cite{Baillard2007}. However, as the accuracy is achieved over a limited spectral range, we transfer it to the blue spectral region, performing an absorption spectroscopy of the 461~nm $\gstate\rightarrow\OnePOne$ transition on the ultracold gas. The spectrum is fit to a Lorentzian profile to determine the resonance frequency, which is found to be 146(5)~MHz above the value reported in the literature~\cite{Florian2014}. This variation, which we attribute to a systematic drift of the wavemeter, is subtracted from each measurement of the 477~nm laser wavelength to ensure a frequency accuracy equal to the accuracy of the wavemeter around the clock transition. 

\begin{figure}
    \centering    \includegraphics[width=\linewidth]{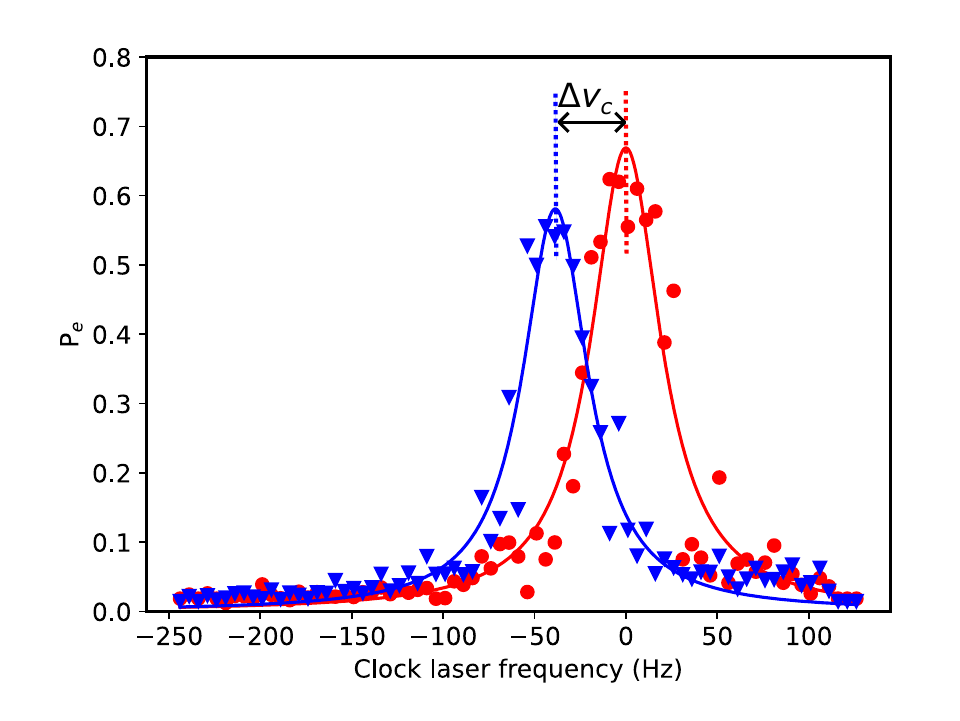}
    \caption{Clock spectroscopy with test laser off (red circles) and test laser on (blue triangles), when the frequency of the 477~nm test laser is set at $f_{\textrm{test}}=628737.85(3)\,$GHz. The excited state population fraction $P_e$ is plotted against the frequency of the clock laser, and fits with a Lorentzian profile (plain curve). The reference frequency is set to be the Lorentzian line center when the test laser is turned off. The difference in frequency between the two line centers gives the clock transition line shift, $\Delta\nu_c$, in the presence of the test 477~nm laser. Here, the test laser intensity is set to $I_0=99(1)~
    \text{W/cm}^2$. Each point is averaged over five measurements. }
   \label{fig:spectrum}
\end{figure}

The timing sequence for each experimental run is shown in~\fig{expt}(b). Initially, $^{88}$Sr atoms are held in the optical lattice for about 150~ms. Then, we turn on a bias coil pair in Helmholtz condition to generate the magnetic field required to activate the clock transition. After a wait of 30~ms for the current in the bias coil to stabilize, we turn on the clock laser for 20~ms to drive the clock transition over a $\pi$-pulse. After the probing stage, the population in the $\gstate$ ground state is measured by fluorescence imaging on the 461~nm transition. The population in the $\ThreePZero$ clock state is measured by optically repumping the atoms back to the ground state by shining the 707~nm and 679~nm lasers. We then capture a second fluorescence image using the 461~nm transition. The experimental sequence is repeated by varying the frequency of the clock laser in steps of 5~Hz around the expected resonance, while turning on and off the 477~nm test laser in an interleaved manner. \fig{spectrum} shows the typical spectra for the two cases at a laser optical frequency of $f_{\textrm{test}}=628737.85(3)\,$GHz.

In the presence of the 477~nm test laser, we expect energy shifts of $\Delta E_{g}$ and $\Delta E_{e}$ on the ground and excited states, respectively [see the inset of \fig{expt}(a)]. If $\Delta E_{g} \neq \Delta E_{e}$, the atom experiences a non-zero clock transition frequency shift of $\Delta \nu_c=(\Delta E_{e}-\Delta E_{g})/h$. To measure this shift in the clock transition frequency, we fit the resulting spectra with a Lorentzian profile to determine the respective line centers. The experimental clock frequency shift $\Delta\nu_c$ is calculated from the frequency difference between the two line centers.
 
\section{Results}\label{sec:results}
\begin{figure}
    \centering    \includegraphics[width=\linewidth]{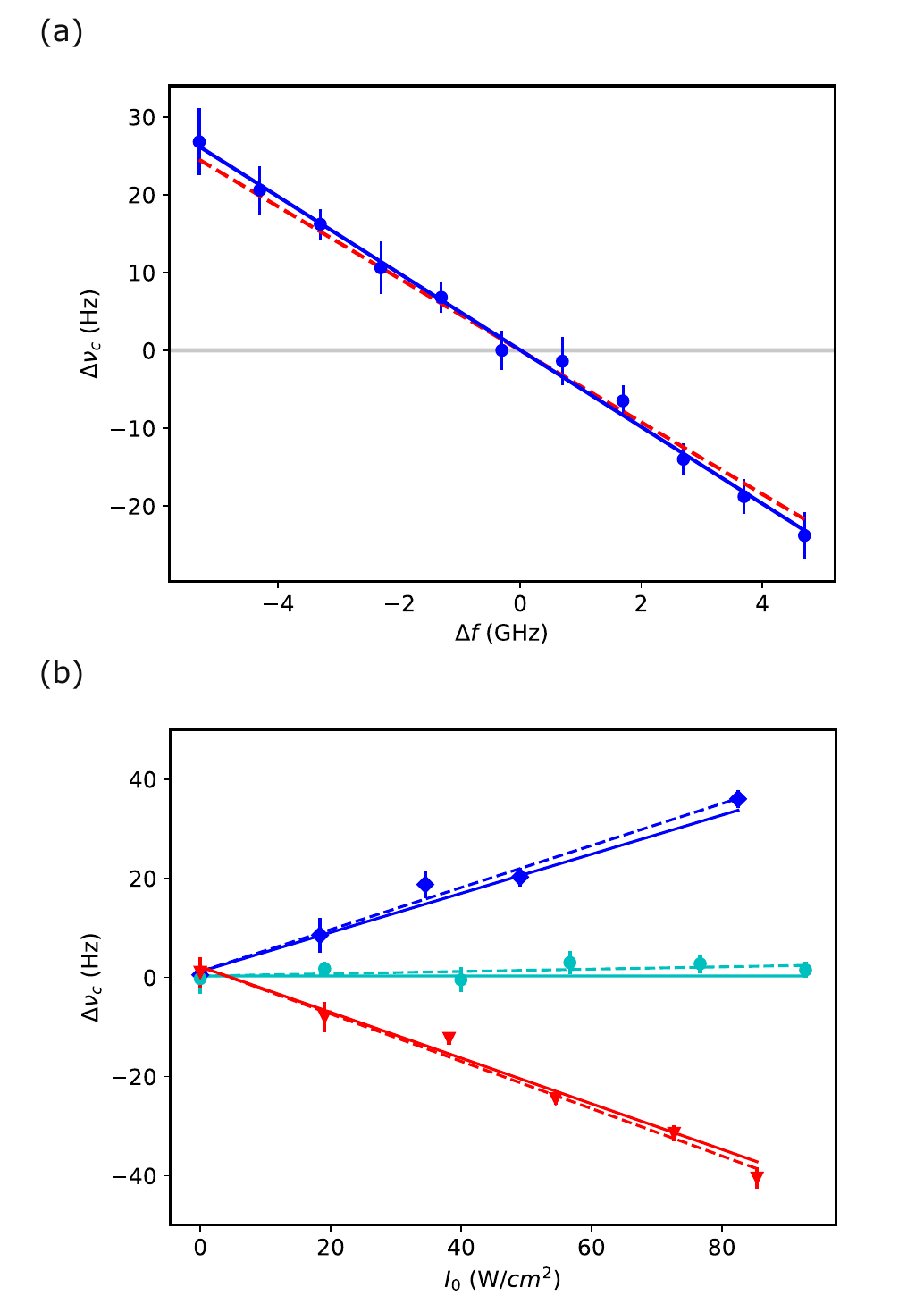}
    \caption{(a) Clock transition shift as a function of frequency detuning from the magic wavelength $\Delta f$. The data points are fit with linear functions depicted as the solid lines. The theoretical prediction, given by \eq{ave_clockshift}, is plotted as the dashed line, taking the experimental value for the magic wavelength. (b) Clock transition shift dependence on the test laser peak intensity for $\Delta f = -8.3(1)$~GHz (blue diamonds), $\Delta f = 0.0(1)$~GHz (cyan circles), and $+9.7(1)$~GHz (red triangle). The dashed and solid lines correspond to the theoretical prediction and linear fitting of experimental data, respectively. The error bars are the standard deviations calculated with five measurements, corresponding to the statistical errors in the experiment (see main text for more details).
    }
    \label{fig:results}
\end{figure}

We start with a coarse scan of the test laser frequency $f_{\textrm{test}}$  for a rough determination of $f_{\textrm{magic}}$ corresponding to the magic wavelength condition, namely $\Delta\nu_c=0$. Then, we vary the test laser frequency over a range of 10~GHz around $f_{\textrm{magic}}$ for a finer estimation. The test laser power is set at $P_0=42.4(3)$~mW, corresponding to a peak intensity of $I_0=99(1)$~W/cm$^2$. The values of $\Delta\nu_c$ for each $f_{test}$ are reported in \fig{results}(a). The experimental results are fit to a linear function, with the inverse of the error bars acting as weights. This fit is shown as the blue solid line in \fig{results}(a). The intersection point between $\Delta\nu_c = 0$ and the fitting line indicates the frequency of the magic wavelength, which is found to be 628728.2(1)~GHz, or equivalently, a wavelength of 476.82362(8)~nm in vacuum. The slope is found to be $-4.9(2)$~Hz/GHz. 

Both the linear fit error and the measurement uncertainty of the wavemeter contribute to the overall error of the reported magic wavelength. For the experimental data points in~\fig{results}, the error bars are the fit error of the measured spectrum, where statistical error in the fluorescence imaging is the main source of uncertainty. AC-Stark shifts due to the fluctuations and drift of the clock laser power do not contribute significantly to the error, since we perform our measurement in an interleaved manner. Shot-to-shot fluctuation in the bias magnetic field leads to a root mean square frequency fluctuation of 0.5~Hz, smaller than the error bars of $\Delta\nu_c$. Moreover, we compute the systematic shift of the magic wavelength due to state mixing induced by the bias magnetic field or the 698-nm and 813-nm optical fields. We find a negligible shift below 1~MHz.

We compare the experimental results with the theoretical prediction. The frequency of the clock transition experiences a differential AC-Stark shift given by,
\begin{equation}
   \Delta \nu = \frac{I}{2h\varepsilon_0c}\Delta\alpha,
\end{equation}
where $\Delta \alpha$ denotes the differential polarizability between the clock states. $I$ is the laser intensity, $\varepsilon_0$, and $c$
are the vacuum permittivity, and the speed of light in vacuum, respectively.
Using the $\alpha$ values reported in Sec.~\ref{sec:theo}, we estimate the magic wavelength to be 476.885(54)~nm, which is in reasonable agreement with the experimental value. The difference of 0.061(54)~nm between the two values is almost consistent with zero, lying just outside the error of one standard deviation.

Additionally, we analyze the variation of the differential AC-Stark shift in the vicinity of the magic wavelength. Within the scanning range of the test laser frequency, we approximate the differential polarizability $\Delta\alpha$ to be linearly dependent on the frequency detuning $\Delta f$. Thus, we have
\begin{equation}
   \frac{\Delta \nu_c}{\Delta f} =\left. \frac{I}{2h\varepsilon_0c}\frac{\textrm{d}\Delta\alpha}{\textrm{d}\Delta f}\right\rvert_{\textrm{magic}} , \label{eq:clockshift}
\end{equation}
where $\textrm{d}\Delta\alpha/\textrm{d}\Delta f|_{\textrm{magic}}= 1.328(1)$~a.u./GHz is the frequency gradient of the clock-transition differential polarizability at the magic wavelength. Taking the test laser peak intensity $I_0$, \eq{clockshift} gives $\left.\Delta\nu_c/\Delta f\right|_{I_0}=-6.2(2)$~Hz/GHz, which is significantly higher than the experimental value of -4.9(2)~Hz/GHz. However, the 477~nm beam size is comparable to the size of the atomic cloud, leading to inhomogeneity of the clock-transition AC-Stark shift. We perform a spatial averaging of the AC-Stark shift, weighted by the normalized Gaussian profile of the atomic cloud, and derive the following simple expression,
\begin{equation}
   \overline{\frac{\Delta \nu_c}{\Delta f}} = \frac{w_x}{\sqrt{w_x^2+4\sigma_x^2}}\frac{w_z}{\sqrt{w_z^2+4\sigma_z^2}}\left.\frac{\Delta \nu_c}{\Delta f}\right|_{I_0}. \label{eq:ave_clockshift}
\end{equation}
We find $\overline{\Delta\nu_c/\Delta f}=-4.6(4)$~Hz/GHz, which is now in agreement with the experiment [see also the red-dashed line in \fig{results}(a)].

To test the model validity over an extended parameter range, we measure the frequency shift of the clock line center $\Delta\nu_c$ for several values of test-laser peak intensity $I_0$, and frequencies $f_{\textrm{test}}$. The results are presented in~\fig{results}(b). The plain lines correspond to linear fits, whereas the dashed lines are extracted from \eq{ave_clockshift}.  

\section{Conclusion}\label{sec:conclusion}

We performed a measurement of a blue magic wavelength near 477~nm in $^{88}$Sr atoms with a relative frequency uncertainty at the $10^{-7}$ level. The magic wavelength is found to be 476.82362(8)~nm from a spectroscopic measurement of the clock-transition AC-Stark shift. Our experimental value is in reasonable agreement with the theoretical prediction, with a difference of  0.061(54) nm, mainly dominated by the uncertainty in the theoretical calculation. Thus, this work could be helpful for a more accurate estimation of the atomic polarizability in the blue spectral region, for example, by fine-tuning the reduced dipole moment of some transitions or the core contribution.

The 477-nm magic wavelength for the strontium clock transition has the advantage of a shorter wavelength and higher polarizability, compared to the commonly used 813-nm magic wavelength. This makes the 477-nm magic wavelength a good candidate for Bragg pulses to perform matter-wave interferometry involving both the $\gstate$ and $\ThreePZero$ states, with potential applications for atomic clock interferometry~\cite{Loriani2019,DiPumpo2021} and quantum test of weak equivalence principle in the optical domain~\cite{zych2018quantum,Rosi2017}. However, the photon scattering rate for a given lattice depth is about three orders of magnitude higher at 477~nm than at 813~nm. Hence, the 477-nm magic wavelength is more challenging to use for conventional optical lattice clocks and strongly correlated gases, two applications that generally require a long coherence time. Nevertheless, a denser atomic sample trapped in 477-nm optical tweezers could be useful to study cooperative effects in light emission, since the experimental timescales are generally shorter in these studies.

Our measurements also provide a reference point to locate other wavelengths of interesting state-dependent potential, such as the anti-magic wavelength or the tune-out wavelength~\cite{Safronova2015}. 

\begin{acknowledgments}

The authors thank Li Jianing for her contribution to the development of the experimental apparatus. This work was supported by the CQT/MoE (Grant No. R-710-002-016-271) and the Singapore Ministry of Education Academic Research Fund Tier2 (Grant No. MOE-T2EP50223-0004).

\end{acknowledgments}


%

\end{document}